\documentclass[11pt]{article}

\usepackage{graphicx,wrapfig,lipsum}
\usepackage{geometry}
\usepackage{setspace}
\usepackage{titling}
\usepackage{amsmath, amssymb, amsthm}
\usepackage[dvipsnames]{xcolor}
\begin{document}

% --- Cover Page ---
\newgeometry{margin=0.001in} 
\begin{titlepage}
    \centering

    % Bandeau image (full width)
    \includegraphics[width=\textwidth]{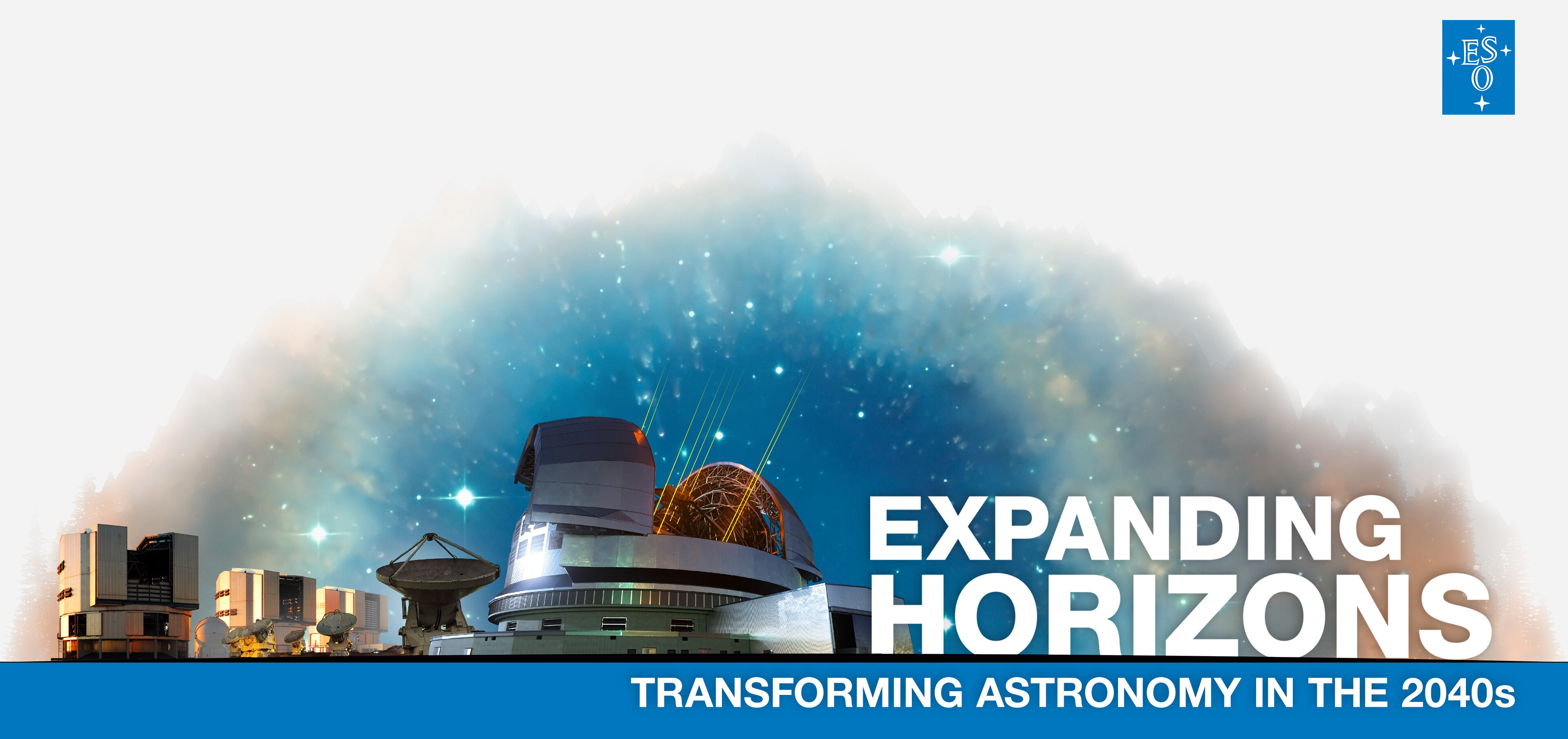}
    \vspace{0.5cm}

    % Title

    {\huge \bfseries Resolving the smallest scales of \\ massive star formation: 
    A case for next-generation \\ thermal-infrared interferometers \par} 
    %Resolving Star Formation Where It Happens: Inner Disks, Fragmentation, and Accretion Flows
    %Resolving the Birth of Massive Stars: A Thermal-Infrared View
    %Precision views of star formation: the need for high-angular and high-spectral resolution"
    \vspace{0.82cm}

    % Authors and affiliations
    {\large
    E. Bordier $^{1}$, E. Koumpia $^{2}$, L. Labadie$^{1}$, J. Sanchez-Bermudez$^{3}$, Á. Sánchez-Monge$^{4}$, J.-P. Berger$^{5}$ }\par
    \vspace{0.75cm}
    {\footnotesize
    $^{1}$ I. Physikalisches Institut, Universitat zu Köln, Zulpicher Str. 77, Cologne, 50937, Germany \\
    $^{2}$ Joint ALMA Observatory / European Southern Observatory, Alonso de C\'ordova 3107, Casilla 19, Santiago, Chile \\
    $^{3}$ Instituto de Astronom\'ia, Universidad Nacional Aut\'onoma de M\'exico, Apdo. Postal 70264, Ciudad de M\'exico, 04510, M\'exico \\
    $^{4}$ Institut de Ciències de l’Espai (ICE), CSIC, Campus UAB, Barcelona, Spain \\
     $^{5}$ Université Grenoble Alpes, CNRS, IPAG, 38000 Grenoble, France \\}
    % }
    % }
    \vspace{1.487cm}
 \includegraphics[width=\textwidth]{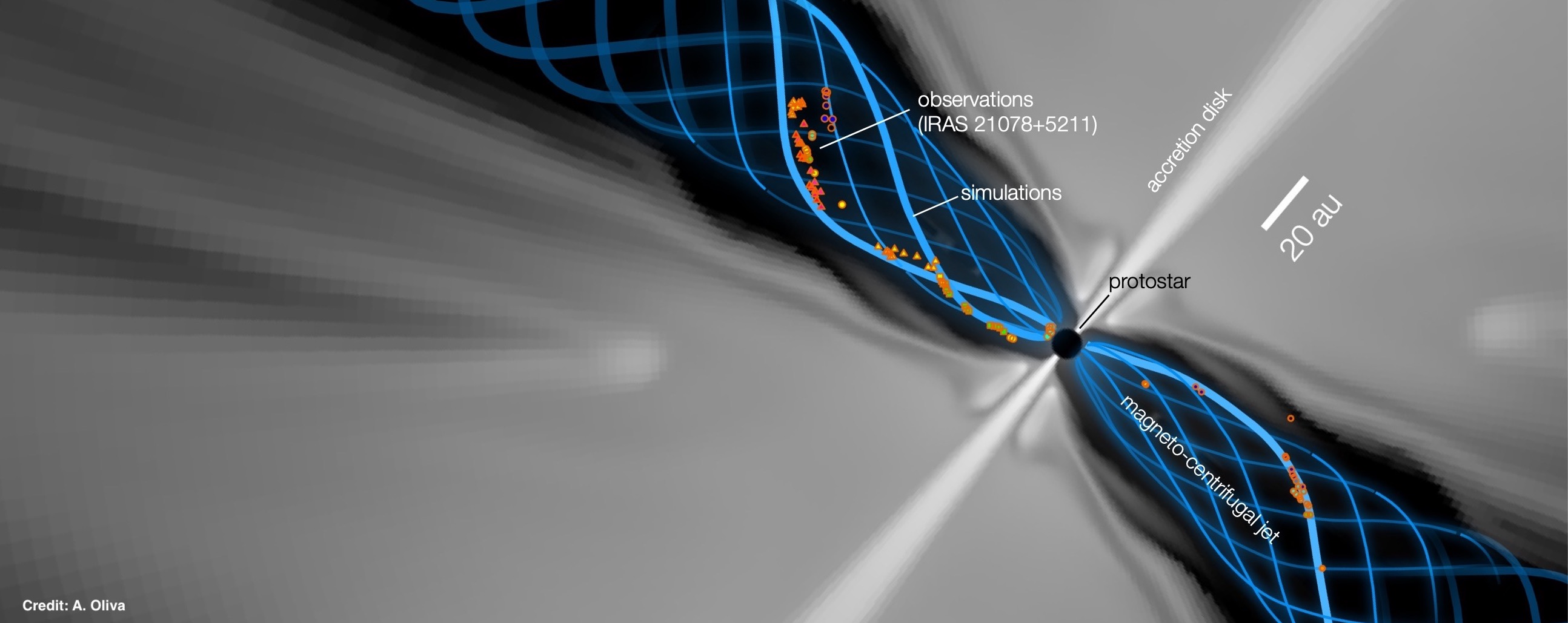}
    \vfill
\end{titlepage}

\newgeometry{margin=0.6in}
\newpage

\textbf{\large \color{RoyalBlue} Abstract:} Understanding how stars assemble their mass across the different environments, from isolated low-mass cores to dense clusters forming massive stars, remains a central challenge in astrophysics (See review by Beuther et al. 2025). Fundamental questions persist about how environmental conditions, turbulence, and metallicity shape star formation, and whether the dominant modes of star formation at different redshifts differ from those observed in the present-day Universe. A critical missing link lies in understanding how large-scale accretion flows that feed parsec-sized gas reservoirs are funneled down to the compact circumstellar disks ($\lesssim$100~au) through which material is ultimately delivered onto forming stars and planetary systems, passing through the crucial star–disk interaction region ($\lesssim$1~au) where accretion, ejection and angular-momentum regulation take place. While ALMA has greatly advanced our view of the larger-scale picture, revealing streamers, bridges, and complex gas flows feeding stellar cores, the smallest and ultimate stages of accretion and ejection processes remain poorly constrained across the entire stellar mass spectrum.
This gap in our understanding is even higher in the case of massive forming stars, given their rarity, embeddedness and high distance at which they are found. These stars require large gas reservoirs and high accretion rates, yet the physical processes governing the innermost few au remain largely inaccessible. In this regime, disks may become gravitationally unstable, fragment into companions, undergo tidal interactions within clusters, and suffer disruption from powerful radiative and mechanical feedback. Whether massive star disks exhibit substructures such as gaps, analogous to planet-forming disks around low-mass stars, remains an open question. Accessing these regions observationally relies on mid-infrared (MIR) dust continuum emission, atomic fine-line structures and molecular vibration–rotation lines that trace ionized gas at 100–1000~K on scales of only a few tens to few au.
% This White Paper outlines the science cases related to massive star formation and especially accretion/ejection processes that require observational capabilities surpassing those of current and planned facilities. 
Addressing these questions demands extreme-angular resolution and high-spectral resolution, which together are essential for resolving the physical processes that govern star formation to the fundamental scales of accretion and disks.

\section{The unresolved physics of massive star formation}

% {\large \color{RoyalBlue} The unresolved physics of massive star formation}
 
  \begin{wrapfigure}{r}{8.8cm}
\includegraphics[width=8.8cm]{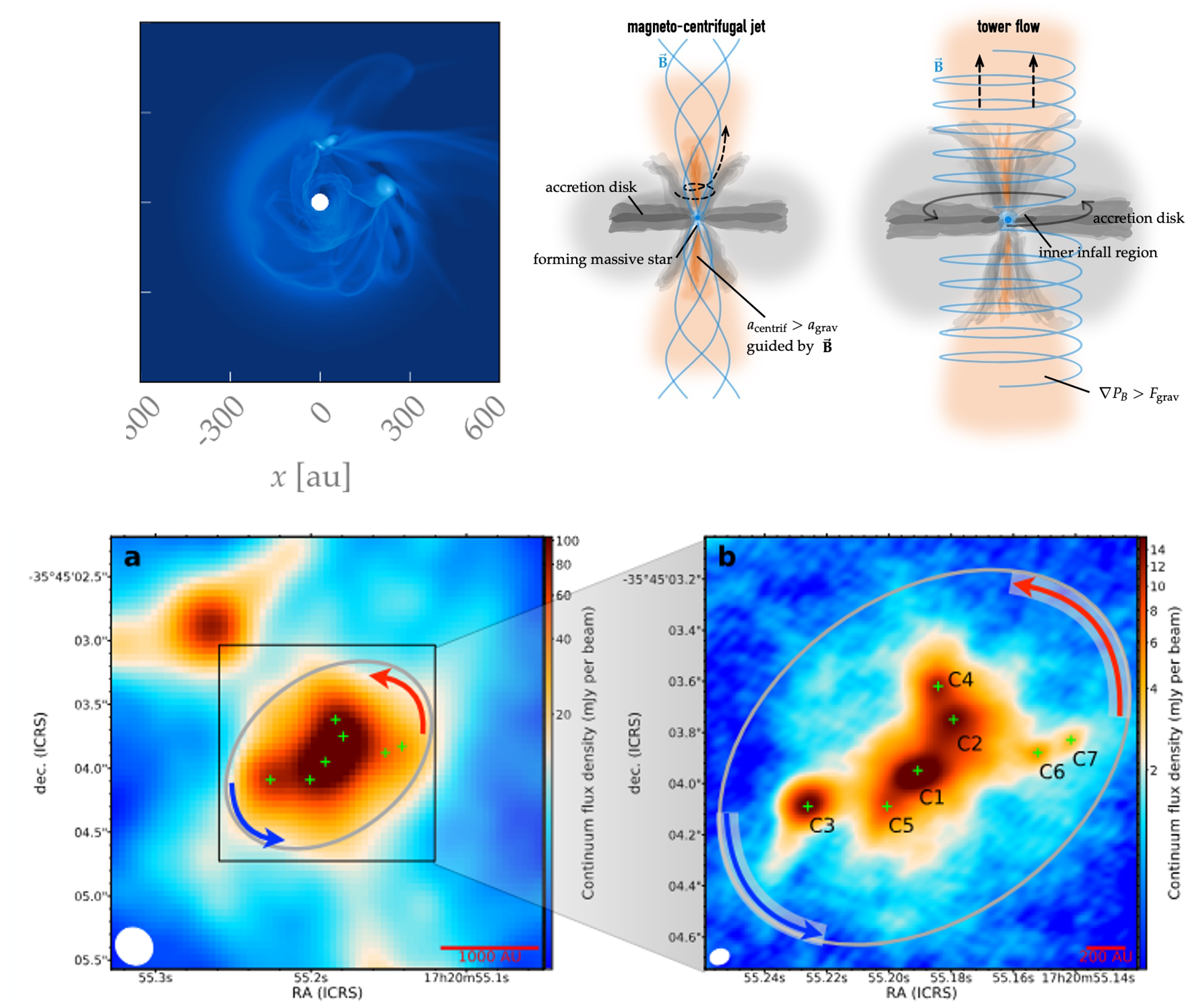}
\caption{\footnotesize \emph{Top}: : PLUTO simulations showing the early stages of disk fragmentation [4] and schematic of massive protostellar outflows in the case of magneto-centrifugal and tower flow mechanisms (Courtesy: A. Oliva). \emph{Bottom}: NGC 6334IN ALMA low- and high-resolution 1.3mm continuum images highlighting that fragmentation is a viable mechanism for the formation of extreme high-order multiplicity [5].}\label{wrap-fig:1}
\end{wrapfigure} 

 Massive stars dominate the energy budget, chemical enrichment, and feedback processes of galaxies, yet their formation pathways remain far less constrained than those of low-mass stars. 
 A leading theoretical framework invokes disk-mediated accretion, in which material is funneled through an equatorial disk while radiation and momentum escape along the poles of the forming star [1,2]. This geometry allows continued growth despite intense radiation pressure and naturally creates conditions for disk fragmentation, potentially leading to high-order multiplicity on 10s$-$100s~au scales [3].

Disks around massive young stellar objects (MYSOs) are short-lived ($<10^{5}$~yr; e.g. [6]), deeply embedded, subject to dispersal by photoevaporative winds, and typically located at distances of several kiloparsecs, making them intrinsically difficult to detect. Theoretical models further suggest that shielding within the innermost $\sim$100 au is essential to overcome the radiation-pressure barrier and enable the formation of stars exceeding $\sim 40~M_{\odot}$ [7]. Resolving these regions requires sub-milliarcsecond (mas) angular resolution, together with sensitivity to warm dust and gas that are opaque at optical and near-infrared (NIR) wavelengths. Current facilities only partially meet these requirements, leaving key questions about \textbf{accretion physics, feedback, and multiplicity} unanswered.

\section{Why the inner 100~au remain elusive}
Over the past decade, ALMA/VLA has revealed rotating gaseous structures and Keplerian disks around a growing number of MYSOs, providing strong evidence that disk-mediated accretion and fragmentation operates at high stellar masses (see e.g.[8-14]). NIR interferometry (VLTI, CHARA) has delivered a small but crucial set of benchmark systems, including resolved disks, disk winds, and compact binaries (see e.g. [16-18]). These few well-characterized objects, now numbering only a handful above $\sim 20~M_{\odot}$, anchor our current understanding.

Despite this progress, major observational gaps remain. Sub-mm observations (ALMA, NOEMA, VLA) are inherently biased toward cooler outer disk regions and cannot probe the star–disk interaction zone or detect key symmetric molecules lacking a dipole moment. NIR interferometric observations, while offering high angular resolution, are strongly affected by extinction (MYSOs emission peak $\sim100~\mu m$) and provide marginal spectral resolution, hence limiting access to the thermal structure and chemistry of the inner disk. As a result, the innermost regions of the youngest high-mass protostellar objects (HMPOs, [19]) and MYSOs where accretion, jet launching, disk fragmentation, and early multiplicity are established remain largely unresolved.
% Moreover, high-mass protostellar objects (HMPOs; [20]), being younger and more deeply embedded than classical MYSOs, are prime targets for probing the onset of multiplicity. 
Access to sub-au scales in these systems would directly address \textbf{how mass is transported from disk to star, where and under which conditions massive disks fragment, how accretion and chemistry co-evolve in the inner 0.1–10 au, and how these processes depend on environment}. Looking ahead, advances beyond GRAVITY+ in angular and spectral resolution and wavelength coverage will be required to both drastically expand Galactic HMPO/MYSO samples and enable population-level studies in nearby galaxies, allowing the role of metallicity, radiation field, and clustering to be tested for the first time. 

% detecting protobinary systems separated by only a few astronomical units requires angular resolutions well below one mas at typical Galactic distances. Even with upcoming upgrades such as GRAVITY+, the accessible parameter space remains restricted to relatively evolved or favorable targets. The brief formation timescales of MYSOs, combined with heavy obscuration and clustering, further compound the difficulty. Consequently, it is still unclear whether the observed formation pathways are universal, or how they depend on environmental factors such as metallicity, radiation field, and clustering.

\section{Accessing warm dust and gas at sub-mas scales}
% A long-baseline thermal-infrared interferometer operating from $\sim$3 to at least 27~$\mu m$ would fundamentally transform this field by bridging the spatial and spectral gaps between existing facilities. 

%   \begin{wrapfigure}{l}{8.8cm}
% \includegraphics[width=8.8cm]{spectrum_MIR.png}
% \caption{\footnotesize MIR spectrum with major atomic fine-structure lines and H$_{2}$O rotational lines marked with an asterisk [21].}\label{wrap-spectrum}
% \end{wrapfigure} 

Achieving sub-mas resolution in the $LMN$, and potentially $Q$ bands would allow direct access of the inner disk regions at sub-au scales, at wavelengths sensitive to warm dust, hot gas, and embedded protostars. 
Such a facility would enable direct measurements of dust location, temperature, composition, size and opacity (for e.g. through the 10$\mu m$ silicate feature), which critically determine disk structure and emission properties. It would also provide access to a rich molecular inventory unavailable at sub-mm wavelengths. In addition to the well-known HeI, Br$\gamma$, NaI, CO-bandheads in the $K-$band tracing the warm star-disk accreting interface, complementary atomic fine-structure lines in $NQ-$bands, including [Fe II] at $\sim$17, 24, and 26~$\mu m$, provide sensitive diagnostics of gas temperature, while lines such as [Ne II] and [S I] at 12.8 and 25.2~$\mu m$ trace gas density and excitation in shocked regions [20]. MIR vibration–rotation lines of species such as CH$_{4}$, C$_{2}$H$_{2}$, HCN, NH$_{3}$, OH, H$_{2}$O and HNCO trace photodissociation and the hot inner ($<$20 au) regions where organic chemistry is established (e.g. [20,22]). Many of these molecules are invisible to ALMA, yet are key building blocks of complex chemistry. High spectral resolution to reach few 10s km/s (R $\approx$ 50,000 in the $LMN$-bands and $\approx$ 25,000 in the $Q-$band) is essential to resolve their narrow line widths and to localize their emission within the disk. In addition, spatially resolved spectroscopy would disentangle emission arising from disks, infall streams, and the inner outflow or jet-launching regions thanks to variability studies and direct probes of accretion and ejection processes operating on dynamical timescales. Polycyclic aromatic hydrocarbons (PAHs), observable throughout the $3-18~\mu m$ range, provide powerful diagnostics of UV irradiation, disk geometry, and gas heating. Differential phase measurements could spatially resolve PAH emission out to tens of au, offering insights into disk flaring and feedback from newly ignited massive stars.

\section{Towards a complete picture of massive star formation}
The science enabled by thermal-infrared interferometry extends beyond individual disks. By accessing massive protostars across a wide range of environments (including lower-metallicity regions) it becomes possible to directly test theoretical predictions that disks become more unstable and fragment more readily under enhanced thermal pressure and photo-evaporation. Emerging observational evidence already suggests that disk fragmentation is a viable pathway to extreme multiplicity [5,13]; a next-generation interferometer would provide the statistical samples needed to confirm this mechanism across environments. Together with ALMA upgrades, ELTs, JWST, and GRAVITY+, it would deliver \textbf{a full chromatic picture of massive star formation}, from cold outer envelopes to hot inner accreting and fragmenting disks, and jets. 
% Extending coverage into the Q band, particularly from high, dry sites comparable to the ALMA plateau, would further enhance sensitivity to the warm dust and gas characteristic of high-mass protostellar objects and accreting zero-age main sequence stars.

% By resolving the innermost regions where accretion, feedback, chemistry, and multiplicity intersect, such a facility would address one of the most persistent gaps in astrophysics. In doing so, it would move the field toward a unified theory of massive star formation capable of explaining how the most influential stars in the Universe are born.

\section{Accessing the $Q-$band and ideas of infrastructure}
\begin{figure}
    \centering
    \includegraphics[width=1\linewidth]{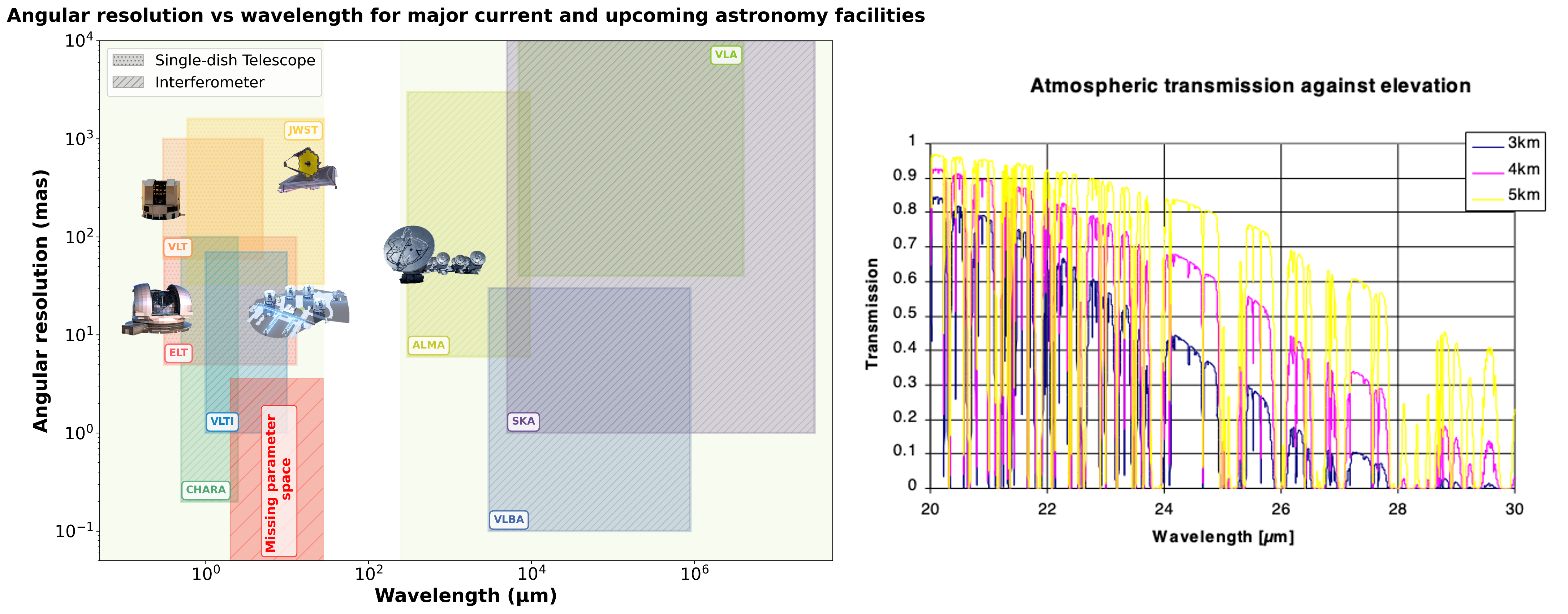}
    \caption{\footnotesize Left: Comparison of spatial resolution of contemporary IR to radio projects, regardless of their sensitivity. Right: Variation of $20-30~\mu m$ atmospheric transmission with elevation of the observing site, using a Zenith mid-latitude winter standard model. From the instrument concept study for T-OWL.}
    \label{fig:placeholder}
\end{figure}

Reaching the Q-band ($\approx$ 20–27~$\mu$m) with sub-mas angular resolution sufficient to probe inner massive-star disks at 10s-100s~kpc distances requires near-km baselines. This presents significant technical challenges, primarily due to the high thermal background and stringent phase stability requirements at these wavelengths. Equally critical is site selection: $Q-$band atmospheric transmission improves dramatically at high, dry plateaus, where low precipitable water vapour reduces absorption and thermal background.
Heterodyne interferometry offers a particularly promising and scalable solution in this regime by converting the incoming infrared signal into a radio-frequency signal at the telescope, enabling coherent combination over km-scale baselines with intrinsically high spectral resolution, independent of thermal background [23,24]. While this approach trades raw continuum sensitivity for spectral resolving power, it is ideally matched to bright, line-rich sources such as massive protostars, evolved stars, and active galactic nuclei, where narrow molecular and atomic features carry key information on kinematics, accretion, and feedback. This naturally enables strong synergies across multiple astrophysical fields.
While large-scale heterodyne infrared interferometry remains technologically challenging, recent laboratory demonstrations and ongoing developments in frequency combs, MIR detectors, and photonic technologies [23,24] indicate a clear and credible development path. A heterodyne-based, long-baseline interferometer deployed at a high, dry site therefore represents the most realistic and transformative pathway toward sub-mas interferometry in the MIR, and more particularly in the 
$N$ and $Q$ bands.

\vspace{0.5cm}

{\color{RoyalBlue} References:}

{\footnotesize 
\textbf{[1]} Kuiper et al. 2015, ApJ 800, 86;
\textbf{[2]} Oliva et al. 2023, A\&A, 669, A80;
\textbf{[3]} Suri et al. 2021, A\&A 655, A84;
\textbf{[4]} Oliva 2020, A\&A  644, A41
\textbf{[5]} Li et al. 2025, NatAst, arxiv:2509.06787
\textbf{[6]} Kuiper, R. \& T. Hosokawa 2018, A\&A, 616, A101 
\textbf{[7]} Tanaka, K. E. I. \& T. Nakamoto 2011, ApJL, 739, L50;
% \textbf{[4]} Kuiper, R. \& T. Hosokawa 2018, A\&A, 616, A101 
% \textbf{[5]} Tanaka, K. E. I. \& T. Nakamoto 2011, ApJL, 739, L50;
\textbf{[8]} Kraus, S. et al. 2010, Nature, 466, 339; 
% Hofmann, K.-H., Menten, K. M.,
\textbf{[9]} Johnston, K. G. et al. 2015, ApJL, 813, L19; 
% Robitaille, T. P., Beuther, H.,
\textbf{[10]} Maud, L. T. et al. 2019, A\&A, 627, L6; %, Cesaroni, R., Kumar, M. S. N.,
\textbf{[11]} Ilee, J. D. et al. 2018, ApJL, 869, L24; %, Cyganowski, C. J., Brogan, C. L.,
\textbf{[12]} S\'anchez-Monge, A. et al. 2013, A\&A, 552, L10; %, Cesaroni, R., Beltr\'an, M. T.,
\textbf{[13]} Li, S. et al. 2024, NatAs, 8, 472L;
% \textbf{[14]} Li, S. et al. 2025, NatAs, arxiv:2509.06787;
\textbf{[14]} S\'anchez-Monge, A. et al. 2025, MNRAS, 543, 662;
\textbf{[15]} Koumpia, E. et al. 2021, A\&A, 654, A109; %, de Wit, W.-J., Kraus, S.,
\textbf{[16]} Koumpia, E. et al. 2024, A\&A, 690, A209;
\textbf{[17]} Bordier, E. et al. 2022, A\&A, 663, A26;
\textbf{[18]} Sanchez-Bermudez, J. et al. 2016, A\&A, 588, A117; %, Hummel, C. A., Tuthill, P.,
\textbf{[19]} Beuther 2007, PPVI, p.165
\textbf{[20]} Tappe, G. et al. 2008, ApJ, 680L, 117T;
\textbf{[21]} Nomura, H. et al 2005, A\&A, 438, 923N;
\textbf{[22]} Ramirez-Tannus, M. C. et al. 2025, A\&A, 701A, 139R;
\textbf{[23]} Bourdarot, G. et al. 2020, A\&A, 639A, 53B;
\textbf{[24]} Allain, T. et al. 2024,SF2A, 179A;

% \textbf{[12]} N\"urnberger, D. E. A., \& Stanke, T. 2003, A\&A, 400, 223; \textbf{[13]} N\"urnberger, D. E. A. 2003, A\&A, 404, 255; 

% \textbf{[15]} Ilee, J. D. et al. 2013, MNRAS, 429, 2960; %, Wheelwright, H. E., Oudmaijer, R. D.,
% \textbf{[16]} Bik, A. A\&A, 427, L13. } %, \& Thi, W.-F. 2004,

%Bordier 2022, A\&A, 663, A26 
%Sanchez-Monge et al. 2025, MNRAS, 543, 662
%Li et al. 2024, NatAst, 8, 472L
%Li et al. 2025, NatAst, arxiv:2509.06787

\end{document}